\documentclass[journal,a4paper]{IEEEtran}
\usepackage[T1]{fontenc}
\usepackage{layouts}

\usepackage{cite}

\usepackage[pdftex,colorlinks=true,bookmarks=true,citecolor=blue,urlcolor=blue]{hyperref} 

\ifCLASSINFOpdf
  \usepackage[pdftex]{graphicx}
   \graphicspath{{./figs/}}
  \DeclareGraphicsExtensions{.pdf,.jpeg,.png}
\else
\fi
\usepackage{amsmath}
\usepackage{amsthm}
\usepackage[cmintegrals]{newtxmath}
\usepackage{bm}

\usepackage{algorithmic}

\usepackage{array}

\ifCLASSOPTIONcompsoc
  \usepackage[caption=false,font=normalsize,labelfont=sf,textfont=sf]{subfig}
\else
  \usepackage[caption=false,font=footnotesize]{subfig}
\fi
\usepackage{mathrsfs}  
\usepackage{amsbsy}

\newcommand{\field}[1]{\mathbb{#1}}
\newcommand{\fs}[1]{\mathsf{#1}}

\DeclareMathOperator{\diag}{diag}

\newcommand{\tp}{\intercal}

\newcommand{\ovl}[1]{\overline{#1}}

\newcommand{\bigO}[1]{\mathop{\mathcal{O}}\left(#1\right)}

\newcommand{\vv}[1]{\mathbf{#1}}
\newcommand{\vs}[1]{\boldsymbol{#1}}

\DeclareMathOperator{\sech}{sech}

\usepackage{enumitem}

\begin{document}
%
\title{Fast Nonlinear Fourier Transform using Chebyshev Polynomials}
%
%
%


\author{Vishal Vaibhav
\thanks{Email:~\tt{vishal.vaibhav@gmail.com}}
}
\maketitle

\begin{abstract}
We explore the class of exponential integrators known as 
exponential time differencing (ETD) method in this letter to design
low complexity nonlinear Fourier transform (NFT) algorithms
that compute discrete approximations of the scattering coefficients 
in terms of the Chebyshev polynomials for real values of the 
spectral parameter. In particular, we discuss ETD Runge-Kutta methods 
which yield algorithms with complexity $O(N\log^2N)$ (where $N$ is
the number of samples of the signal) and an order of convergence 
that matches the underlying one-step method.
\end{abstract}


%
\IEEEpeerreviewmaketitle

\section{Introduction}
This letter addresses the signal processing aspect of a nonlinear Fourier transform 
(NFT) based optical fiber communication system which has emerged as one of the 
potential ways of mitigating nonlinear signal distortions at higher signal 
powers~\cite{Yousefi2014compact,TPLWFK2017}. The idea consists in encoding 
information in the nonlinear Fourier spectrum of the signal which is 
synthesized using the inverse NFT. The information can be decoded using the 
direct NFT which forms the primary focus of this letter. Let us note that 
inverse NFT algorithm of extremely high order of convergence for computing the 
radiative potential have been proposed~\cite{V2019BL1}, which coupled with 
a higher order convergent direct NFT via the Darboux transformation procedure, 
could yield a full inverse NFT with higher order of convergence. 

Recently, integrating factor (IF) based exponential integrators have been successfully 
exploited for the purpose of developing fast direct as well as inverse NFT 
algorithms~\cite{V2017INFT1,V2018BL,V2018LPT,V2019LPT}. In this letter, we introduce 
another class exponential integrators which are known as 
\emph{exponential time differencing} (ETD) method, a name originally used 
in computational electrodynamics, for the purpose of designing fast NFT 
algorithms. This particular class of algorithms rely on FFT based fast 
polynomial arithmetic in the \emph{Chebyshev basis} as opposed to the monomial basis 
considered in the aforementioned works. While ETD schemes do not have any of the 
drawbacks associated with the IF schemes~\cite{CM2002}, the ETD schemes offer 
an additional advantage that there are no restrictions on the kind of the 
grid employed for sampling the potential. All ETD schemes within the Runge-Kutta 
as well as linear multistep methods can be sped up. In this letter, however, we have 
chosen the Runge-Kutta methods for our exposition.

We begin our discussion with a brief review of the scattering theory closely
following the formalism presented in~\cite{V2017INFT1}. The NFT of any signal is 
defined via the Zakharov-Shabat (ZS) scattering 
problem (henceforth referred to as the \emph{ZS problem}) which can be stated as 
follows: For $\zeta\in\field{R}$ and $\vv{v}=(v_1,v_2)^{\tp}$,
$\vv{v}_t =\left[-i\zeta\sigma_3+U(t)\right]\vv{v}$
where $\sigma_3=\diag(1,-1)$. The potential 
$U(t)$ is defined by $U_{11}=U_{22}=0,\,U_{12}=q(t)$ and $U_{21}=r(t)$ with $r=\alpha q^*$
($\alpha\in\{+1, -1\}$). Here, $\zeta\in\field{R}$ is known as the \emph{spectral parameter}
and $q(t)$ is the complex-valued signal. The solution of the ZS problem consists in finding the so called 
\emph{scattering coefficients} which are defined through special solutions 
known as the \emph{Jost solutions} which are linearly 
independent solutions of the ZS problem such that they have a plane-wave 
like behavior at $+\infty$ or $-\infty$. The Jost solution of the 
\emph{second kind}, denoted by $\vs{\phi}(t,\zeta)$, has 
the asymptotic behavior $\vs{\phi}(t;\zeta)e^{i\zeta t}\rightarrow(1,0)^{\tp}$ 
as $t\rightarrow-\infty$. The asymptotic behavior 
$\vs{\phi}(t;\zeta)\rightarrow (a(\zeta)e^{-i\zeta t}, b(\zeta)e^{i\zeta
t})^{\tp}$ as $t\rightarrow\infty$ determines the scattering coefficients $a(\zeta)$ and
$b(\zeta)$ for $\zeta\in\field{R}$. In this letter, we primarily focus on 
the continuous spectrum, also referred to as the 
\emph{reflection coefficient}, which is defined by 
$\rho(\xi)={b(\xi)}/{a(\xi)}$ for $\xi\in\field{R}$. 
\section{The Numerical Scheme}\label{eq:num-scheme}
The exponential time differencing Runge-Kutta (ETD--RK) methods with $s$ 
internal stages can be characterized much in the same way as the standard 
Runge-Kutta method adapted to the ZS problem: Given the step-size $h>0$, an ordered set of nodes 
$c_j\in[0,1],\,j=1,2,\ldots,s$ defining the abscissas 
$t_n\leq t_n+c_1h\leq t_n+c_2h\leq\ldots\leq t_n+c_sh\leq t_{n+1}$ and the 
matrix-valued weight functions $a_{ij}(-i\zeta h\sigma_3)$, 
$b_{i}(-i\zeta h\sigma_3)$, the internal stages (for $j=1,2,\ldots,s$) of the ETD--RK 
method together with the final update read as
\begin{equation}
\begin{split}
&\vv{v}_{n,j}=e^{-ic_j\zeta h\sigma_3}\vv{v}_{n} 
+ h\sum_{k=1}^sa_{jk}(-i\zeta h\sigma_3){U}_{n+c_k}{\vv{v}}_{n,k},\\
&\vv{v}_{n+1}
=e^{-i\zeta h\sigma_3}\vv{v}_{n}
+h\sum_{k=1}^sb_{k}(-i\zeta h\sigma_3){U}_{n+c_k}\vv{v}_{n,k},
\end{split}
\end{equation}
where we have used the convention ${U}_{n+c_k}=U(t_n+c_kh)$. The linear 
system stated above can be written as
\begin{equation}\label{eq:linear-sys-ETD-RK}
\begin{pmatrix}
I_{2s\times 2s}-A(\zeta;h)D_n & 0_{2s\times2}\\
-B(\zeta;h)^{\tp}D_n & \sigma_0
\end{pmatrix}
\begin{pmatrix}
{\vs{\Upsilon}}_n\\
{\vv{v}}_{n+1}
\end{pmatrix}=
\begin{pmatrix}
E(\zeta;h){\vv{v}}_{n}\\
e^{-i\zeta h \sigma_3}{\vv{v}}_{n}\\
\end{pmatrix},
\end{equation}
where ${D}_n=\diag\left(h{U}_{n+c_1},h{U}_{n+c_1},\ldots,h{U}_{n+c_s}\right)$, 
$A(\zeta;h)= \left(a_{ij}(-i\zeta h\sigma_3)\right)\in\field{C}^{2s\times 2s}$, and, 
$B(\zeta;h)$, $E(\zeta;h)\in\field{C}^{2s\times 2}$ and 
$\vs{\Upsilon}_n\in\field{C}^{2s}$ are defined as
\begin{equation}
\begin{split}
& B= \begin{pmatrix}
b_{1}(-i\zeta h\sigma_3)\\
\vdots\\
b_{s}(-i\zeta h\sigma_3)
\end{pmatrix},\,\,
E=
\begin{pmatrix}
e^{-ic_1\zeta h\sigma_3}\\
\vdots\\
e^{-ic_s\zeta h\sigma_3}
\end{pmatrix},\,\,
{\vs{\Upsilon}}_n=
\begin{pmatrix}
{\vv{v}}_{n,1}\\
\vdots\\
{\vv{v}}_{n,s}
\end{pmatrix}.
\end{split}
\end{equation}
The existence of a solution of the linear system~\eqref{eq:linear-sys-ETD-RK} depends on the 
determinant $\Delta_{n+1}(\zeta;h)=\det\left[I_{2s\times 2s}-A(\zeta;h)D_n\right]$
which can be shown to be nonzero for sufficiently small $h$. For explicit RK 
methods, $\Delta\equiv1$. Introducing the transfer 
matrix ${M}_{n+1}(\zeta;h)$, the numerical scheme can be stated as
\begin{equation}\label{eq:ETD-RK-TM}
\vv{v}_{n+1}={e^{-i\zeta h}}[\Delta_{n+1}]^{-1}{M}_{n+1}(\zeta; h)\vv{v}_n.
\end{equation} 
Let us remark that collocation method~\cite{HO2005} with Legendre--Gauss--Lobatto nodes is one of 
the simplest methods of constructing implicit ETD--RK methods. The weight
functions in this case can be stated as follows: For $k=1,2,\ldots,s$, we have $a_{1k}=0$ and 
\begin{equation}
\begin{split}
&a_{jk}(z)=\int_0^{c_j}e^{(c_j-\tau)z}L_k(\tau)d\tau,\quad j=2,3,\ldots,s-1,\\
&b_k(z)=\int_0^{1}e^{(1-\tau)z}L_k(\tau)d\tau=a_{sk}(z),\\
&\text{where}\,\,L_k(\tau)
=\prod_{n\neq k,n=1}^s\frac{\tau-c_n}{c_k-c_n}
=\sum_{n=1}^s\frac{\lambda_{n-1}^{(k)}}{(n-1)!}\tau^{n-1}.
\end{split}
\end{equation}
Evidently, $a_{jk}(z)=\sum_{n=1}^s{c_j^{n}\lambda_{n-1}^{(k)}}\varphi_{n}(c_jz)$ where 
\begin{equation}\label{eq:vphi-def}
(n-1)!\varphi_n(z) = \int_0^1 e^{(1-\tau)z}{\tau^{n-1}}d\tau.
\end{equation}
The $\varphi$-functions defined above are central to all ETD schemes; therefore, 
we discuss it in more detail in Sec.~\ref{app:varphi_func}. In the following, we use the 
convention $\varphi_0(z)=e^{z}$, 
$\varphi_{j,k}=\varphi_j(-ic_k\zeta h\sigma_3)$ and $\varphi_{j}=\varphi_j(-i\zeta h\sigma_3)$
in order to represent the ETD--RK methods using the Butcher Tableau. An example of an 
implicit three-stage ETD--RK method is the Lobatto~IIIA method of order $4$, labelled as 
ETD--IRK$_{34}$, which is given by
\begin{equation*}
\renewcommand\arraystretch{1.2}
\begin{array}
{c|ccc}
0   & 0 & 0 &0\\
\frac{1}{2} &
\frac{1}{2}\varphi_{3,2}-\frac{3}{4}\varphi_{2,2}+\frac{1}{2}\varphi_{1,2} 
& \varphi_{2,2}-\varphi_{3,2}
&\frac{1}{2}\varphi_{3,2}-\frac{1}{4}\varphi_{2,2}\\
1   &4\varphi_3-3\varphi_2+\varphi_1 
    &4\varphi_2-8\varphi_3
    &4\varphi_3-\varphi_2\\
\hline
  &4\varphi_3-3\varphi_2+\varphi_1
  &4\varphi_2-8\varphi_3
  &4\varphi_3-\varphi_2
\end{array}
\end{equation*}
Noting that $\vv{v}_{n+1}=\vv{v}_{n,3}$, the transfer matrix relation can be written 
as $\vv{v}_{n+1}=e^{-i\zeta h}\Xi_1^{-1}\Xi_2\vv{v}_{n}$ where
\begin{equation*}
\left\{\begin{aligned}
&\Xi_1=\left[\sigma_0-a_{33}(-i\zeta h\sigma_3)U_{n+1}\right]
\det\Gamma_{n+1/2}\\
&\quad-a_{32}(-i\zeta h\sigma_3)U_{n+1/2}\Gamma_{n+1/2}
a_{23}(-i\zeta h\sigma_3)U_{n+1},\\
&\Xi_2=\left[e^{-i\zeta h\sigma_3}+a_{31}(-i\zeta h\sigma_3)U_{n}\right]\det\Gamma_{n+1/2}\\
&+ a_{32}(-i\zeta h\sigma_3)U_{n+1/2}\Gamma_{n+1/2}
\left[e^{-\frac{1}{2}i\zeta h\sigma_3}+a_{21}(-i\zeta h\sigma_3)U_{n}\right],
\end{aligned}\right.
\end{equation*}
with $\Gamma_{n+1/2}=\sigma_0+a_{22}(-i\zeta h\sigma_3)U_{n+1/2}$. Example of an 
explicit four-stage ETD--RK method of order $4$ is given by~\cite{CM2002}
\begin{equation*}
\renewcommand\arraystretch{1.1}
\begin{array}
{c|cccc}
0   & 0 & 0 &0&0\\
\frac{1}{2} & \frac{1}{2}\varphi_{1,2} &0&0&0\\
\frac{1}{2} &0& \frac{1}{2}\varphi_{1,2} &0&0\\
1   & \frac{1}{2}\varphi_{1,2}(\varphi_{0,2}-\sigma_0) & 0 &\varphi_{1,2}&0\\
\hline
  &4\varphi_3-3\varphi_2+\varphi_1 & 2\varphi_2-4\varphi_3 &2\varphi_2-4\varphi_3 & 4\varphi_3-\varphi_2
\end{array}
\end{equation*}
which we label as ETD--ERK$_{34}$.
\subsection{Fast algorithm for Chebyshev nodes}
Let the computational domain be $\Omega=[T_1,T_2]$ and set $2T=T_2-T_1$. Let the number
of steps be $N_s$ and $h=2T/N_s$. The grid is defined by $t_n= T_1 + nh/2,\,\,n=0,1,\ldots,N,$ with 
$t_{N}=T_2$ where $N=2N_s$ is the number of samples. Consider the Jost solution 
$\vs{\phi}(t;\zeta)$: $\vs{\phi}(T_1;\zeta)=(1,0)^{\tp}e^{-i\zeta T_1}$ and 
${\vs{\phi}}(T_2;\zeta)=(a(\zeta)e^{-i\zeta T_2},b(\zeta)e^{+i\zeta T_2})^{\tp}$. Let the 
cumulative products be denoted by 
$\mathcal{M}_{1\to N_{s}}={M_{N_{s}}\times\ldots\times M_{1}}$ and
$\mathcal{D}_{1\to N_{s}}
={\Delta_{N_{s}}\times\ldots\times \Delta_{1}}$
so that the discrete approximation to the scattering 
coefficients, $a$ and $b$, can be worked out to be 
$a_N = {\left(\mathcal{M}_{1\to N_{s}}\right)_{11}}/\mathcal{D}_{1\to N_{s}}$ and 
$b_N = e^{-i2\zeta T_2}\left(\mathcal{M}_{1\to N_{s}}\right)_{21}/\mathcal{D}_{1\to N_{s}}$.

Now, let us consider the problem of computing the samples of the continuous spectrum 
at the $N$ Chebyshev-Gauss-Lobatto (CGL) nodes given by 
$\xi_j=-(\pi/2h)\cos\left[{\pi j}/(N-1)\right]$ for 
$j=0,\ldots,N-1$. Let us first note that the matrix-valued weight 
functions can be computed 
beforehand. The naive approach to forming the cumulative products 
$\mathcal{M}_{1\to N_{s}}$ and $\mathcal{D}_{1\to N_{s}}$ by direct 
multiplication evidently yields a complexity of $\bigO{N_sN}=\bigO{N^2}$. The 
fast algorithm, on the other hand, is based on the idea that each of the transfer matrices 
$M_n(\zeta;h)$ and the determinants $\Delta_n(\zeta;h)$ can be represented 
using a small number of Chebyshev polynomials. The motivation behind this step 
is to use fast polynomial arithmetic in Chebyshev basis for computing the discrete 
scattering coefficients. To this end, define
\begin{equation}
M_{n\to n+m-1}=M_{n+m-1}\times M_{n+m-2}\times\ldots\times M_{n},
\end{equation} 
where $m$ is a power of $2$. The strategy is to evaluate 
$M_{n\to n+m-1}(\xi;h)$ on $d+1$ CGL nodes by direct 
multiplication of the constituting matrices followed by a discrete Chebyshev 
transform to obtain the Chebyshev expansion coefficients:
\begin{equation}\label{eq:TM-cheb-expand}
M_{n\to n+m-1}(\xi;h)=\sum_{j=0}^{d}M^{(j)}_{n\to n+m-1}
\mathcal{T}_j\left(\frac{2}{\pi}\xi h\right),
\end{equation} 
for $\xi\in[-\pi/2h,\pi/2h]$ where $\mathcal{T}_j(\cdot)$ denotes the Chebyshev 
polynomial of degree $j$. Next, we also extend the strategy to the 
determinants $\Delta_n$.
If the total number of 
matrices (or determinants) to be multiplied in the Chebyshev basis was 
$N_s$ originally, the aforementioned step reduces this number to $N_s/m$ 
at an initial cost of $9md + 5m^{-1}N_s\nu(d+1)$ where $\nu(n)$ is the 
cost of discrete Chebyshev transform of size $n$.

The next step is to compute the products of the matrices (or determinants) 
pairwise as in~\cite{V2018LPT} except we use FCT instead of FFT.
Let $\nu(n)$, where $n$ is a power of $2$, be the complexity of multiplying two
polynomials of degree $n-1$ in the Chebyshev basis. Let $\ovl{d}=d+1$ be a power of $2$ 
and $\varpi(n)$ denote the complexity of multiplying $n$ matrices, then 
$\varpi(n)=8\nu(\ovl{d}n/2)+2\varpi(n/2)$. The overall complexity now depends on 
how the polynomial products are carried out in the Chebyshev basis which is 
considered next. Let us note that the Chebyshev transform of a complex vector 
of size $d+1$ is equivalent to the DFT of a vector of size $2d$; therefore, the 
complexity of the fast Chebyshev algorithm is $\varpi_{\text{FFT}}(2d)$~\cite{G2011} 
where $\varpi_{\text{FFT}}(n)$ denotes the complexity of FFT of size $n$.

In the following, we describe a fast algorithm for multiplying polynomials in the 
Chebyshev basis which proceeds by defining an equivalent multiplication problem in the monomial 
basis~\cite{G2011}: Let us consider the polynomials $p(x)$ and $q(x)$ of 
degree $D=N-1$ stated in the Chebyshev basis: 
$p(x) = {\hat{p}_0}/{2}+\sum_{j=1}^{D}\hat{p}_j\mathcal{T}_j(x)$ and 
$q(x)={\hat{q}_0}/{2}+\sum_{j=1}^{D}\hat{q}_j\mathcal{T}_j(x)$. Let $r(x)=p(x)q(x)$ be expressed as 
$r(x)={\hat{r}_0}/{2}+\sum_{j=0}^{2D}\hat{r}_j\mathcal{T}_j(x)$.
Define $\alpha(x)=\sum_{j=1}^{D}p_jx^j$, $\beta(x)= \sum_{j=1}^{D}q_jx^j$ and 
$\bar{\alpha}(x) = \alpha(x^{-1})x^D$. Also define 
$\gamma(x)=[p_0+\alpha(x)][q_0+\beta(x)]= \sum_{j=0}^{2D}\gamma_jx^j$ and 
$\delta(x)=\bar{\alpha}(x)\beta(x)=\sum_{j=0}^{2D}\delta_jx^j$, then it can be shown by direct 
multiplication and using the property 
$2\mathcal{T}_i(x)\mathcal{T}_j(x)=\mathcal{T}_{i+j}(x)+\mathcal{T}_{|i-j|}(x)$, 
for all $i,j\in\field{Z}_+,$ that
\begin{equation}
2r_j=
\left\{\begin{aligned}
&\gamma_0+2\delta_D, &&j=0,\\
&\gamma_j+\delta_{D-j} + \delta_{D+j}, &&j=1,\ldots,D-1,\\
&\gamma_j, &&j=D,\ldots,2D.
\end{aligned}\right.
\end{equation}
Let us show that the algorithm discussed above can be implemented 
using $4$ FFTs as opposed to $6$ FFTs of size $2N$ leading to a complexity 
estimate of $4\varpi_{\text{FFT}}(2N)+8N$. The computation of $\gamma(x)$ is straightforward 
yielding a complexity of $3\varpi_{\text{FFT}}(2N)+4N$. The computation of $\delta(x)$ can be 
carried out by employing no more than $1$ FFT of size $2N$ as follows: Define 
$\mu(x)=x\delta(x)=x\bar{\alpha}(x)\beta(x)$ whose degree 
is $2D+1<2N$. Putting $\omega=e^{i2\pi/2N}$ and $\omega_k=\omega^k$, we have
$\mu_k = \mu(\omega_k)=\omega\cdot \omega^{Dk}\alpha(\omega^{-k})\beta(\omega^{k})
=(-1)^{k}\alpha(\omega^{2N-k})\beta(\omega^{k})$
which can be easily computed using the quantities involved in computing $\gamma(x)$. 
Finally, the inverse FFT of the sequence $\mu_k$ followed by a shift is required in order 
to determine $\delta(x)$ which costs $\varpi_{\text{FFT}}(2N)+4N$ multiplications 
yielding an overall complexity of $4\varpi_{\text{FFT}}(2N)$ where we have retained 
only the leading term.

Noting that $\varpi_{\text{FFT}}(n)=\bigO{n\log_2 n}$, the asymptotic complexity 
of computing the scattering coefficients using the polynomial multiplication algorithm
described above works out to be 
$\varpi(N_s)=\bigO{m^{-1}\ovl{d}N_s\log^2N_s}$. The original problem of evaluating 
the continuous spectrum at $N$ CGL nodes now reduces to carrying out an inverse discrete 
Chebyshev transform.

\subsection{The exponential functions in ETD}\label{app:varphi_func}
In ETD--RK methods, the weights turn out to be linear combinations of the so-called 
$\varphi$--functions defined by~\eqref{eq:vphi-def}. The first few of 
these functions are $\varphi_1=(e^z-1)/{z}$, $\varphi_2=(e^z-z-1)/{z^2}$ and 
$\varphi_3=(e^z-{z^2}/{2}-z-1)/{z^3}$. These functions have a 
removable singularity at $z=0$ and it can be easily shown
that $\varphi_k(0)={1}/{k!}$. The higher-order $\varphi$-functions can be 
obtained from the recurrence relation 
$z\varphi_k(z)={\varphi_{k-1}(z)-{1}/{(k-1)!}}$ which is well-conditioned for 
$|z|\geq1$. For $|z|<1$, the computation of $\varphi$-functions using the 
expressions above suffers from cancellation errors~\cite{BSW2007}; therefore, we 
resort to their Pad\'e approximants for $|z|<\epsilon$ ($\epsilon=2^{-8}$). The 
$[d/d]$--Pad\'e approximant~\cite{BSW2007} is given by
$\varphi_{\ell}(z)={N_d^{(\ell)}(z)}/{D_d^{(\ell)}(z)}+\bigO{z^{2d+1}}$ with
\begin{equation}
N_d^{(\ell)}=\sum_{j=0}^d\left(\sum_{k=0}^j\nu_{j,k}\right)z^j=\sum_{j=0}^d\nu_{j}z^j,\quad
D_d^{(\ell)}=\sum_{j=0}^d\delta_jz^j,
\end{equation}
where the coefficients can be computed using the recurrence relations 
\begin{equation}
\begin{split}
\nu_{j,k+1} &=
-\frac{(d-k)(\ell+j-k)}{(k+1)(2d+\ell-k)}\nu_{j,k},\quad
\nu_{j,0}=\frac{1}{(\ell+j)!},\\
\delta_{j+1} &= \frac{(d-j)}{(j+1)(2d+\ell-j)}\delta_{j},\quad
\delta_{0}=1.
\end{split}
\end{equation}
For $|z|<1$, we first scale the variable $z$ to $z2^{-K}$, where $K$ is such 
that $|z|2^{-K}<\epsilon$, and use the $[7/7]$--Pad\'e approximants 
stated above. Once the values of $\varphi_{j}(z2^{-K})$ are 
available, $\varphi_{j}(z)$ can be computed by a scaling
procedure given by~\cite{BSW2007}
\begin{equation*}
\begin{split}
&2^{2k}\varphi_{2k}(2z)=\varphi_{k}(z)\varphi_{k}(z)
+\sum_{j=k+1}^{2k}\frac{2\varphi_j(z)}{(2k-j)!},\\
&2^{2k+1}\varphi_{2k+1}(2z)=
\left[\varphi_{k}(z)+\frac{1}{k!}\right]\varphi_{k+1}(z)
+\sum_{j=k+2}^{2k+1}\frac{2\varphi_j(z)}{(2k+1-j)!}.
\end{split}
\end{equation*}
The first few of these scaling relations are as follows:
$2\varphi_1(2z)=(e^z+1)\varphi_1(z)$,
$4\varphi_2(2z)=\varphi_1(z)\varphi_1(z)+2\varphi_2(z)$ and 
$8\varphi_3(2z)=\varphi_1(z)\varphi_2(z)+2\varphi_3(z)+\varphi_2(z)$.

\begin{figure}[!ht]
\begin{center}
\includegraphics[scale=1]{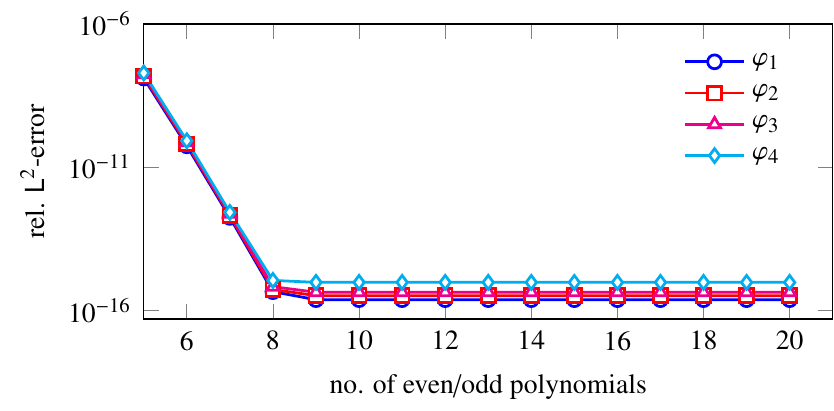}%
\caption{\label{fig:convg-vphi} The figure shows the relative $\fs{L}^2$-error in approximating 
$\varphi_k$ by $\varphi_k(i\xi;2N)$ as a function of $N$ over $[-\pi/2,\pi/2]$.}
\end{center}
\end{figure}

\begin{figure*}[!ht]
\begin{center}
\includegraphics[scale=1]{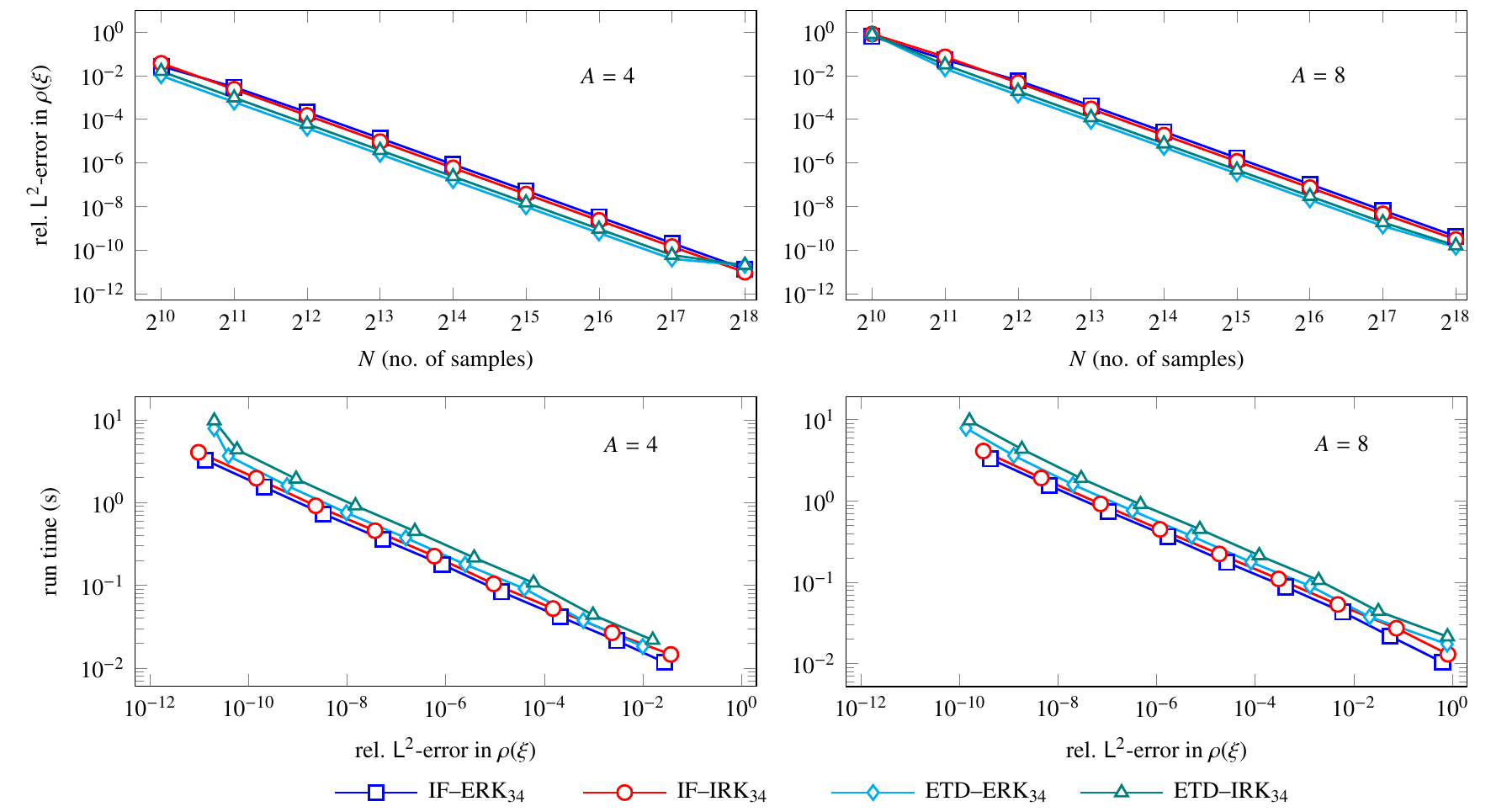}%
\caption{\label{fig:results} The convergence behavior of various NFT algorithms for the chirped 
secant-hyperbolic profile is plotted in the top row. The plots in the
bottom row show the accuracy-complexity trade-off for each of the aforementioned algorithms.}
\end{center}
\end{figure*}

\subsection{Polynomial approximation}\label{app:vphi-poly}
For $z=i\xi$ with $\xi\in[-\sigma,\sigma]$, it is possible to develop 
polynomial approximations similar to that proposed in~\cite{S2014} 
for the $\varphi_k$ functions which converge 
exponentially. This can serve as an alternative to the rational 
Pad\'e approximation method presented above. In addition, this 
furnishes a direct proof of the fact that the 
approximation~\eqref{eq:TM-cheb-expand} is exponentially accurate. Employing 
the expansion
$\exp[i(1-\tau)\xi] =
\sum_{n=0}^{\infty}(2n+1)i^nj_n(\sigma)L_n\left[{(1-\tau)\xi}/\sigma\right]$
where $L_n(\cdot)$ denotes the Legendre polynomial 
of degree $n$ and $j_n(\cdot)$ denotes the spherical Bessel function of order $n$, we have
\begin{equation}\label{eq:varphi-series}
\begin{split}
\varphi_k(i\xi) 
&=\sum_{n=0}^{\infty}(2n+1)i^nj_n(\sigma)
\int_0^1\frac{\tau^{k-1}L_n\left(\frac{1-\tau}{\sigma}\xi\right)}{(k-1)!}d\tau\\
&=\sum_{n=0}^{\infty}(2n+1)i^nj_n(\sigma)P_n\left(\frac{\xi}{\sigma};k\right),
\quad k\geq1.
\end{split}
\end{equation}
The first few polynomials work out to be: $P_{0}\left(x;k\right)={1}/{k!}$,
$P_{1}\left(x;k\right)={x}/{(k+1)!}$ and 
$P_{2}\left(x;k\right)={3x^2}/{(k+2)!}-{1}/(2\,k!)$.
For $m\geq1$, we have the recurrence relations
\begin{equation}
\begin{split}
&P_{2m}
=2x\frac{\left(2m-\frac{1}{2}\right)}{2m+k}P_{2m-1}
-\frac{(2m-k-1)}{2m+k}P_{2m-2}\\
&\qquad+\frac{(-1)^m\left(2m-\frac{1}{2}\right)\Gamma\left(m-\frac{1}{2}\right)}
{\sqrt{\pi}(2m+k)\Gamma(m+1)\Gamma(k)},\\
&P_{2m+1}
=2x\frac{\left(2m+\frac{1}{2}\right)}{2m+1+k}P_{2m}
-\frac{(2m-k)}{2m+1+k}P_{2m-1},
\end{split}
\end{equation}
where we have suppressed the dependence on $x$ and $k$ for the sake of 
brevity. 

Now, let $\varphi_k(i\xi;N)$ denote the approximation using the first $N$ terms of the 
expansion in~\eqref{eq:varphi-series} and 
$\epsilon_N=|\varphi_k(i\xi)-\varphi_k(i\xi;N)|$ the corresponding error. From the inequalities
$|P_n\left(x;k\right)|\leq{1}/{k!}$ and 
$|(2n+1)j_n(\sigma)|\leq e\sqrt{{\sigma}/{2}}\left[{e\sigma}/{(2n+3)}\right]^n$, and, setting 
$\rho_N=(2N+3)/e\sigma>1$ (i.e. assuming $2N>e\sigma-3$), we have
$\epsilon_N\leq\sum_{n=N}^{\infty}{|(2n+1)j_n(\sigma)|}/{k!}\leq C_N\rho_N^{-N}$
where $C_N=\sqrt{{e^2\sigma}/{2}}\rho_N/[(\rho_N-1)k!]$ which confirms that the rate of convergence is 
exponential with respect to $N$ (see Fig.~\ref{fig:convg-vphi} for a numerical example). 
\subsection{Numerical tests}\label{sec:num-exp}
For the numerical experiments, we employ the chirped secant-hyperbolic 
potential given by 
$q(t) = A\sech(t)\exp[2iA\log\sech(t)]$ ($\alpha=-1$) considered 
in~\cite{V2019LPT}. We set the computational domain to be $[-30,30]$ and 
let $A\in\{4,8\}$. Let $\Omega_h=[-\pi/2h,\pi/2h]$ denote the $\xi$-domain; then, the
error in computing $\rho(\xi)$ is quantified by  
$e_{\text{rel.}}=
\|\rho(\xi)-\rho_{\text{num.}}(\xi)\|_{\fs{L}^2(\Omega_h)}
/\|\rho(\xi)\|_{\fs{L}^2(\Omega_h)}$
where $\rho_{\text{num.}}(\xi)$ is the numerical approximation and the integrals are 
computed using the trapezoidal rule. For the purpose of testing, we include 
the RK methods presented in~\cite{V2019LPT} which are labelled as IF--ERK$_{34}$ 
and IF--IRK$_{34}$. The results are plotted in Fig.~\ref{fig:results} which comprises 
convergence analysis (top row) and the trade-off between accuracy and complexity
(bottom row). The results show that the ETD methods tend to be more accurate that the 
IF methods but their complexity-accuracy trade-off is comparable. A superior multiplication 
algorithm for polynomials in the Chebyshev basis may, however, 
dramatically change this situation.

\bibliographystyle{IEEEtran}

\begin{thebibliography}{10}
\providecommand{\url}[1]{#1}
\csname url@samestyle\endcsname
\providecommand{\newblock}{\relax}
\providecommand{\bibinfo}[2]{#2}
\providecommand{\BIBentrySTDinterwordspacing}{\spaceskip=0pt\relax}
\providecommand{\BIBentryALTinterwordstretchfactor}{4}
\providecommand{\BIBentryALTinterwordspacing}{\spaceskip=\fontdimen2\font plus
\BIBentryALTinterwordstretchfactor\fontdimen3\font minus
  \fontdimen4\font\relax}
\providecommand{\BIBforeignlanguage}[2]{{%
\expandafter\ifx\csname l@#1\endcsname\relax
\typeout{** WARNING: IEEEtran.bst: No hyphenation pattern has been}%
\typeout{** loaded for the language `#1'. Using the pattern for}%
\typeout{** the default language instead.}%
\else
\language=\csname l@#1\endcsname
\fi
#2}}
\providecommand{\BIBdecl}{\relax}
\BIBdecl

\bibitem{Yousefi2014compact}
M.~I. Yousefi and F.~R. Kschischang, ``Information transmission using the
  nonlinear {F}ourier transform, {P}art {I},'' \emph{IEEE Trans. Inf. Theory},
  vol.~60, no.~7, pp. 4312--4369, 2014.

\bibitem{TPLWFK2017}
S.~K. Turitsyn, J.~E. Prilepsky, S.~T. Le, S.~Wahls, L.~L. Frumin, M.~Kamalian,
  and S.~A. Derevyanko, ``Nonlinear {F}ourier transform for optical data
  processing and transmission: advances and perspectives,'' \emph{Optica},
  vol.~4, no.~3, pp. 307--322, Mar 2017.

\bibitem{V2019BL1}
V.~Vaibhav, ``Nonlinearly bandlimited signals,'' \emph{J. Phys. A: Math.
  Theor.}, vol.~52, no.~10, p. 105202, 2019.

\bibitem{V2017INFT1}
------, ``Fast inverse nonlinear {F}ourier transformation using exponential
  one-step methods: {D}arboux transformation,'' \emph{Phys. Rev. E}, vol.~96,
  p. 063302, 2017.

\bibitem{V2018BL}
------, ``Fast inverse nonlinear {F}ourier transform,'' \emph{Phys. Rev. E},
  vol.~98, p. 013304, 2018.

\bibitem{V2018LPT}
------, ``Higher order convergent fast nonlinear {F}ourier transform,''
  \emph{IEEE Photonics Technol. Lett.}, vol.~30, no.~8, pp. 700--703, 2018.

\bibitem{V2019LPT}
------, ``Efficient nonlinear {F}ourier transform algorithms of order four on
  equispaced grid,'' \emph{IEEE Photonics Technol. Lett.}, vol.~31, no.~15, pp.
  1269--1272, 2019.

\bibitem{CM2002}
S.~M. Cox and P.~C. Matthews, ``Exponential time differencing for stiff
  systems,'' \emph{J. Comput. Phys.}, vol. 176, no.~2, pp. 430--455, 2002.

\bibitem{HO2005}
M.~Hochbruck and A.~Ostermann, ``Exponential {R}unge--{K}utta methods for
  parabolic problems,'' \emph{Appl. Numer. Math.}, vol.~53, no.~2, pp.
  323--339, 2005.

\bibitem{G2011}
P.~Giorgi, ``On polynomial multiplication in {C}hebyshev basis,'' \emph{IEEE
  Trans. Comput.}, vol.~61, no.~6, pp. 780--789, 2011.

\bibitem{BSW2007}
H.~Berland, B.~Skaflestad, and W.~M. Wright, ``{EXPINT}---{A} {MATLAB} package
  for exponential integrators,'' \emph{ACM Trans. Math. Softw.}, vol.~33,
  no.~1, 2007.

\bibitem{S2014}
A.~Y. Suhov, ``An accurate polynomial approximation of exponential
  integrators,'' \emph{J. Sci. Comput.}, vol.~60, no.~3, pp. 684--698, 2014.

\end{thebibliography}

\providecommand{\noopsort}[1]{}\providecommand{\singleletter}[1]{#1}%

\end{document}